# Diffraction properties of lights with transverse orbital angular momentum

Shunlin Huang,[1, 2, 3,4] Peng Wang,[1] Xiong Shen,[1] Jun Liu,[1, 3,*] and Ruxin Li[1, 3]

[1]*State Key Laboratory of High Field Laser Physics and CAS Center for Excellence in Ultra-intense Laser Science, Shanghai Institute of Optics and Fine Mechanics, Chinese Academy of Sciences, Shanghai 201800, China*
[2]*School of Physics Science and Engineering, Tongji University, Shanghai 200092, China*
[3]*University Center of Materials Science and Optoelectronics Engineering, University of Chinese Academy of Sciences, Beijing 100049, China*
[4] *e-mail: huangshunlin@126.com*
[*] *Corresponding author: jliu@siom.ac.cn*

**Abstract**
Spatiotemporal optical vortex (STOV) is a unique optical vortex with phase singularity in the space-time domain and the photons in a STOV can carry transverse orbital angular momentum (OAM). The STOV shows many fantastic properties which are worth exploring. Here, we theoretically and experimentally study the diffraction property of STOV, which is a fundamental wave phenomenon. The diffraction behaviors of STOVs are obviously affected by the transverse OAM. The diffraction patterns of STOV pulses diffracted by a grating show multi-lobe structure with each gap corresponding to 1 topological charge. The diffraction properties of lights with transverse OAM are demonstrated clearly and help us understanding the physical properties of STOV, which will be of special applications, such as the realization of fast detection of STOVs with different topological charges, which may pay the way for STOV based optical communication.

Photons carrying orbital angular momentum (OAM) is an intrinsic property of optical vortex beams. The orientations of the OAM can be parallel or orthogonal to the propagation direction of the light beam, which are classified as longitudinal and transverse OAM, respectively. Conventional spatial optical vortex beam with phase singularity in the spatial plane possesses longitudinal OAM had been demonstrated [1]. Optical vortex beams have wide applications, such as optical tweezer and microparticle manipulation[2, 3], stimulated emission depletion (STED) microscopy[4], optical communication [5]. Spatiotemporal optical vortex (STOV) is a novel optical vortex with phase singularity in the space-time domain and the photons in a STOV can own transverse orbital angular momentum (OAM) [6-8]. Recently, STOVs were observed in experiments [9] and can be generated by using 4 *f* pulse shaper system [10, 11]. Furthermore, significant properties of STOV are analyzed, such as the conservation of transverse OAM in nonlinear optical process, second harmonic generation [12, 13], the propagation and generation properties of STOV [10, 14, 15], angular momenta and spin-orbit interaction of STOV [16] and transverse shifts and time delays occurred when STOV reflected and refracted at a planar interface [17]. Moreover, technologies for the measurement of STOV have also been presented, such as transient-grating single-shot supercontinuum spectral interferometry (TG-SSSI) [18] and interference methods [11]. The STOV beam opens a new area of vortex beam and may has special applications. Understanding the physical properties of STOV is not only important in theoretical areas, but also meaningful for the practical application of STOV.

  Diffraction is a fundamental phenomenon of light wave, which is well known for a conventional polychrome light beam. However, the diffraction property of STOV light has not been reported, to our knowledge. As it is well known that, for a conventional light beam without spatiotemporal coupling, the spectra are diffracted into one continuous line. However, for a STOV light beam with phase singularity in the space-time domain, energy is coupled between space domain and time domain, which may affect the spectral intensity distribution in the wavepacket of STOV. The transverse OAM makes the fundamental optical phenomena, such as diffraction, unique in contrast with conventional light beam. Here, we theoretically and experimentally analyze the diffraction properties of STOV pulses generated using a 4 $f$ pulse shaper system. Interesting, the diffraction patterns have multi-lobe structures. For a STOV with topological charge $l$ = n, the diffraction pattern has n + 1 lobes. And there are n gaps in the diffraction structure, each gap corresponding to $2\pi$ phase winding or topological charge $l$ = 1. By using this diffraction property, one camera together with a grating can be used to directly and fastly recognize STOV pulses with different topological charges.

The STOV pulses are generated using a conventional 4 $f$ pulse shaper system, where the distances between the gratings (1200 line/mm, blaze wavelength 750 nm), the cylindrical lenses ($f$ = 300 mm) and the phase mask are the same and equal to the focal length of the cylindrical lens. And then, the output pulses after the second grating is focused using a spherical lens ($f$ = 1 m), the STOV pulses are obtained in the focal plane [10]. A grating (the third grating) with the same optical parameters as those used in the STOV generator is set in the focal plane of the spherical lens, which is used to diffract the generated STOV pulses. Another cylindrical lens ($f$ = 300 mm) is placed 300 mm away from the third grating, and a camera is placed in the focal plane of this cylindrical lens, which is marked as the observed plane (OP), to achieve the diffraction patterns of the STOVs. Femtosecond pulses with broad spectral bandwidth from a mode-locked Ti:sapphire laser are used as laser pulse source for the 4 $f$ pulse shaper system. The spectra of the femtosecond pulses can own different spectral bandwidths tailored using different spectral filters.

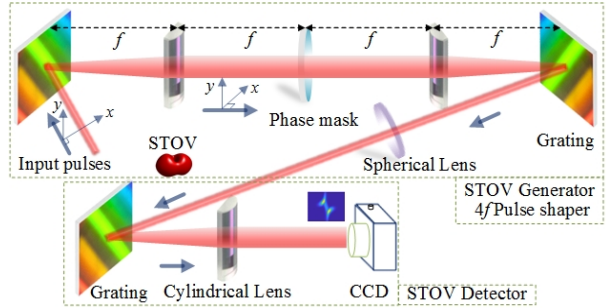

Fig. 1. Experimental setup for the generation of STOV pulses (top part) and for the detection of the diffraction patterns of STOV pulses (bottom part). All the cylindrical lenses own the same focal length $f$ = 300 mm. Fused silica phase masks are inserted at the Fourier plane of the cylindrical lens. The input laser pulses are come from a mode-locked Ti:sapphire laser.

The intensity profiles and the diffraction patterns of the ideal STOV pulses in the $x$-$y$ plane are theoretically analyzed here. For simplicity, the field of the pulse generated from the mode-locked Ti:sapphire laser is described as a Gaussian pulse which has both Gaussian profiles in the spatial and temporal domains with waist radii of $a$ and $b,$ respectively, as

$$E_0(x_1, y_1, t) = \exp[-(x_1^2 + y_1^2)/a^2]\exp[-t^2/b^2], \qquad (1)$$

the Fourier-transform-limited pulse duration of the Gaussian pulse in the calculation is about 200 fs. Then, the wave packet of an ideal STOV pulse with topological charge $l$ ($l\neq0$) can be expressed as [10]

$$E_1(x_1, y_1, t) = [t/b + i(l/|l|)y_1/a]^{|l|} E_0(x_1, y_1, t), \qquad (2)$$

when $l = 0$, the STOV pulse turns into a Gaussian pulse in both the space and time domain. The field of the STOV in the spectrum domain can be obtained by Fourier transform of $E_1(x_1, y_1, t)$ as

$$E_1(x_1, y_1, w) = \int_{-\infty}^{+\infty} E_1(x_1, y_1, t) \exp(-iwt) dt,  \qquad (3)$$

after the STOV is diffracted by a grating, the first order field immediately after the grating can be expressed as [19]

$$E_2(x_2, y_2, w) = E_1(\beta x_2, y_2, w) \exp[i\gamma(w - w_c)x_2], \qquad (4)$$

where $\gamma = 2\pi / w_c d \cos(\theta_{d0})$, $\beta = \cos(\theta_{i0}) / \cos(\theta_{d0})$, $\theta_{d0}$ and $\theta_{i0}$ are the diffracted angle and incident angle of the central frequency ($w_c$) ray, respectively. $w_c = 2\pi c / \lambda_c$, where $c$ is the vacuum light velocity, $\lambda_c$ is the center wavelength. In the simulation and in the experiment, $\lambda_c$ is set at 760 nm. Then $E_2(x_2, y_2, w)$ propagates to the cylindrical lens with a distance of $z = f$, $f$ is the focal length of the cylindrical lens. The field before the cylindrical lens can be written as

$$E_3(x_3, y_3, w) = IFT\{FT\{E_2(x_2, y_2, w)\} \cdot H(f_x, f_y, w)\}, \qquad (5)$$

where $H(f_x, f_y, w) = \exp[ikz - i\pi\lambda z(f_x^2 + f_y^2)]$, and $k$ is the wave vector, $IFT$ and $FT$ denote the spatial inverse Fourier transform and the Fourier transform. $f_x$, $f_y$ are spatial frequencies. The wave packet after passing through the cylindrical lens can be described as

$$E_4(x_4, y_4, w) = E_3(x_3, y_3, w) \exp[-ikx_3^2 / (2f)], \qquad (6)$$

and then the electrical field continuously propagates to the observed plane (OP), the field in OP is marked as $E_5(x_5, y_5, w)$. The spatial intensity profile of the STOV pulse in the focal plane of the spherical lens, and the diffraction pattern of STOV in the OP can be obtained as

$$I(x, y, w) = |E(x, y, w)|^2. \qquad (7)$$

To show the generality of the diffraction property of STOV pulses, the intensity profiles and the diffraction patterns of STOVs with $l = \pm 1$ and $l = \pm 4$ are taken for example. The results are shown in Fig. 2, diffraction patterns of STOVs with $l = \pm 1, \pm 2, \pm 3, \pm 4$ can be seen in the supplement 1. The intensity profiles of the ideal STOVs with $l = \pm 1$ and $l = \pm 4$ in the x-y plane are shown in the first row of Fig. 2, respectively, and the corresponding diffraction patterns are shown in the third row of Fig. 2. As it can be seen from row 1 of Fig. 2, the intensity profiles of the STOVs are not Gaussian spatial distribution, which look like two balls overlap with each other, where the gap between two balls is increased with the increasing of the topological charge. As shown in row 3 of Fig. 2, for the ideal STOV pulses with $l = n$, the diffraction patterns show special multi-lobe structures with n + 1 lobes and n gaps. The dislocated lobes are parallel to each other. And the two lobes in the two end possess higher energies which is more obvious for higher order STOV. The gap number is consistent with the topological charge, hence one gap corresponding to 1 topological charge or $2\pi$ phase winding of the STOV. Moreover, the helicity of the STOV can be obtained clearly from the orientation of the multi-lobe structure.

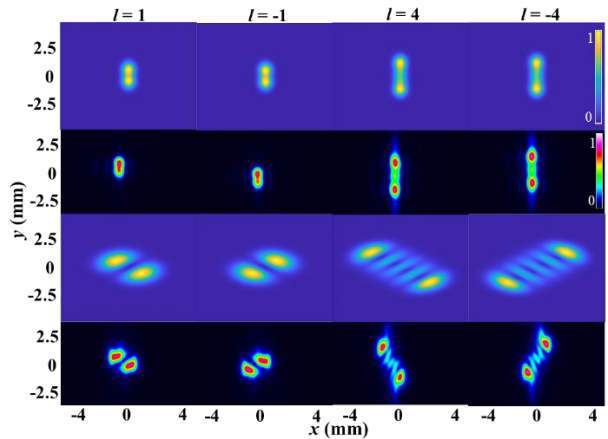

Fig. 2. Intensity profiles and the diffraction patterns of STOVs ($l = \pm1, \pm4$). Row 1 and 3 show the simulated intensity profiles of ideal STOV pulses and their corresponding diffraction patterns, respectively. Row 2 and 4 show the measured results, which are captured by the CCD directly. The intensity profiles and the diffraction patterns of the STOV pulses are calculated or measured in the focal plane of the spherical lens and in the observed plane, respectively.

The measured results of the spatial intensity profiles of the STOVs in the focal plane of the spherical lens and their corresponding diffraction patterns are shown in the second row and the fourth row of Fig. 2, respectively. For the spatial intensity profiles of STOVs, the features of two balls overlapped with different separated distances are shown clearly. And the diffraction patterns of STOVs with different topological charges also show multi-lobe structures. It indicates the diffraction rule for STOVs with different topological charges is consistent with the simulated result well. The helicity of the STOV can also be identified from the orientation of the multi-lobe structure. All the obtained key feature structures and properties from the experimental results, such as the multi-lobe structures, the gap numbers, and the two energetic head lobes match the simulated results very well. It has to be noted that, the generation of STOV is based on space-to-time mapping, where the phase mask is inserted in the Fourier plane ($x$-$y$ plane) to realize the generation of STOV in the space-time plane ($y$-$t$ plane). Then, the generation of the diffraction pattern of STOV can be seen as time-to-space mapping, where the phase circulation in the space-time domain ($y$-$t$ plane) is mapped to the intensity distribution in the space domain ($x$-$y$ plane). Hence, the diffraction is a decoding process. It means that the value of the topological charge can be obtained directly without any retrieval algorithm, which enable the fast recognition of STOVs with different topological charges. Photos of the diffraction patterns captured by the CCD are directly used here to show this fast and clear detection property. The fast recognition of STOVs may pay the way for the wide applications of the STOV pulses, such as STOV based optical communication. This method for the detection of STOVs can be called diffraction method (D-method).

The formation of the unique diffraction pattern of STOV beam, can be attributed to the transverse OAM affected spectral intensity distribution (TOASID) in the wavepacket of STOV. The simulated three dimensional (3D) isosurface plots of the intensity profiles of the STOVs with $l = \pm1, \pm4$ in the $x$-$y$-$w$ are shown in the first column of Fig. 3, respectively. The isosurface is plot at 15% of the max intensity. Then the 3D isosurfaces are viewed in the three main planes, namely isosurfaces viewed in the $x$-$y$ plane, $y$-$w$ plane and $x$-$w$ plane are shown in column 2, 3 and 4, respectively. As it can be seen clearly from row 1 that, the spectral intensity distributions of STOVs show multi-lobe structures. When the 3D STOVs projected to the $x$-$y$ planes, lobes overlapped with each other, and it results in the generation of intensity profiles of STOVs shown in the first row of Fig. 2. Column 3 and column 4 show that the lobes with different spectra actually separated in the $y$ direction and they are parallel to each other and the two head lobes of the structures possess more energies. After diffraction, different lobes of STOV diffracted to different locations in *x* direction and dislocated in *y* direction and then the unique diffraction patterns are obtained.

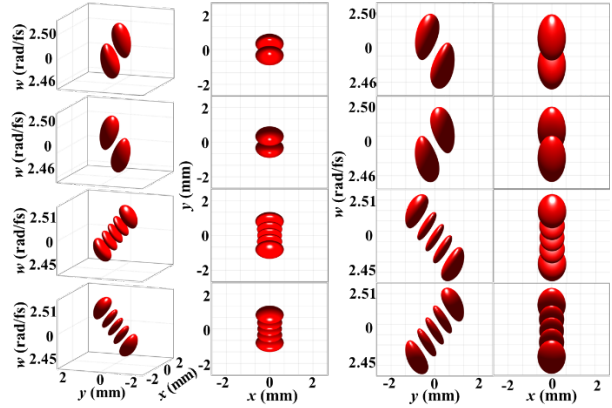

Fig. 3. Column 1 show the simulated three dimensional (3D) isosurface plots of the STOV pulses with *l* = 1, -1, 4, -4 (from row 1 to row 4) in the x-y-w domain, respectively. And they are viewed in the *x-y*, *y-w*, *x-w* planes shown in column 2, 3, 4, respectively.

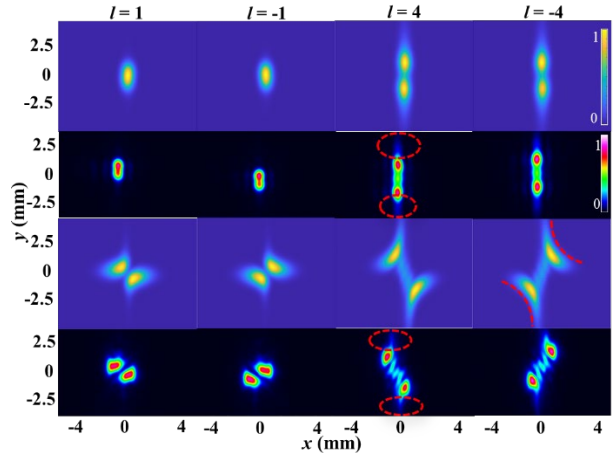

Fig. 4. Intensity profiles of STOVs (*l* = ±1, ±4) and the diffraction patterns. Row 1 and 3 show the simulated intensity profiles of STOV pulses generated from 4 *f* pulse shaper, and their corresponding diffraction patterns, respectively. Row 2 and 4 show the corresponding experimental results.

From the measured results shown in Fig. 2, it can be seen that, weak trails occurred on the two energetic heads in both the intensity profiles of the STOVs and the diffraction patterns. These properties are shown again in row 2 and row 4 of Fig.4 which are marked with red circles or dash lines. The trails mainly originate from the STOV pulses generated from the 4 *f* pulse shaper that are not ideal ones, since the phase mask in the Fourier plane do not match the beam profile in this plane well [10]. As it has been demonstrated that, STOV pulses generated from 4 *f* pulse shaper may have special spatiotemporal intensity profiles and deviate from ideal STOV pulses, such as elliptic intensity profiles with non-uniform energy distribution and multi-hole structures for higher order SOTVs [10, 11, 14], which can be seen from Fig. S1 in the supplement. This is verified by using the STOV pulses generated from the 4 *f* pulse shaper system in the simulation, the method is the same with our early work shown in reference [14]. The simulated intensity profiles of the SOTVs and the corresponding diffraction patterns are shown in the first row and third row of Fig. 4, respectively. Weak trails occur both in the simulated intensity profiles and the corresponding simulated diffraction patterns of the SOTVs. The simulated results match the experimental results well. It can also be seen that the two head lobes in the diffraction patterns show a little bit curve, such as marked with two red dash curve lines shown in column 4 row 3 in Fig. 4, which is another feature of non-ideal STOV. This feature will be more obvious for STOV with broader spectra as shown in Fig. 5.

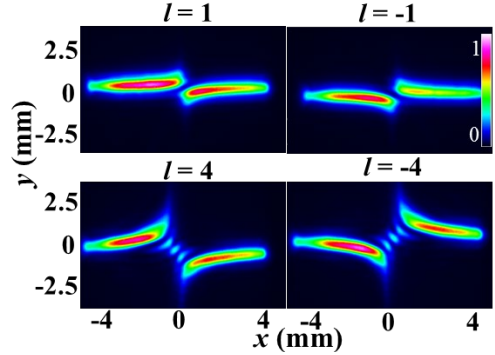

Fig. 5. Experimental diffraction patterns of STOV ($l = \pm 1, \pm 4$) pulses with a broad spectral bandwidth of about 20 nm generated from 4 $f$ pulse shaper.

To verify the diffraction rule is still suitable for STOVs with broader spectral bandwidth which corresponding to shorter pulse duration. Then STOV pulses with a full width at half max (FWHM) spectral bandwidth of about 20 nm are diffracted using the same experimental setup. The diffraction patterns of the STOVs with $l = \pm 1, \pm 4$ are shown in Fig. 5. The key features of the diffraction patterns of STOVs are demonstrated clearly. Moreover, the two head lobes contain wide spectra and spread to a larger area in the $x$ direction and show a little bit curve the same as the simulated results shown in row 3 in Fig. 4. The rule of the generation of lobes and gaps are consistent with the forgoing simulated and experimental results for STOVs with a narrow spectral bandwidth. Furthermore, STOVs with different topological charges can still be recognized well, which shows that the proposed D-method is suitable for the fast recognition of STOVs with different spectral bandwidth, namely STOV pulses with different pulse durations can be detected.

The diffraction properties of lights with transvers OAM are demonstrated theoretically and experimentally. The rules of the diffraction of STOVs are presented. The unique key features of the diffraction patterns of STOV pulses with different spectral bandwidths originate from the TOASID effect. The experimental results match the simulated results well. The diffraction properties of STOV beams help us understanding the physical properties of STOVs, and may have special applications, such as for fast detection of STOVs with different topological charges, which is required in the future wide applications of STOVs such as STOV based optical communication.

**Funding.** National Natural Science Foundation of China (NSFC) (61527821, 61521093, 61905257, U1930115), Chinese Academy of Sciences (CAS) (the Strategic Priority Research Program (XDB160106)), Shanghai Municipal Natural Science Foundation of China (20ZR1464500).

# Diffraction properties of lights with transverse orbital angular momentum: supplemental document

Shunlin Huang,[1, 2, 3,4] Peng Wang,[1] Xiong Shen,[1] Jun Liu,[1, 3,*] and Ruxin Li[1, 3]

[1]*State Key Laboratory of High Field Laser Physics and CAS Center for Excellence in Ultra-intense Laser Science, Shanghai Institute of Optics and Fine Mechanics, Chinese Academy of Sciences, Shanghai 201800, China*
[2]*School of Physics Science and Engineering, Tongji University, Shanghai 200092, China*
[3]*University Center of Materials Science and Optoelectronics Engineering, University of Chinese Academy of Sciences, Beijing 100049, China*
[4] *e-mail: huangshunlin@126.com*
*\* Corresponding author: jliu@siom.ac.cn*

## 1. Experimental setup

Laser pulses from a mode-locked Ti:sapphire laser (Vitara, Coherent) with the center wavelength tailored to 760 nm and 80 MHz repetition frequency are used as the input pulse source for the 4 $f$ pulse shaper system. STOV pulses with controllable spectral bandwidth are obtained by using different spectral filters. Fused silica phase mask with different topological charges are inserted in the Fourier plane of the pulse shaper. The output pulses from the pulse shaper are focused using a spherical lens with focal length of 1 m to generate STOV pulses in the far field. A third grating (1200 line/mm, central wavelength 750 nm) is located in the focal plane of the spherical lens to diffract the generated STOV pulses. After passing through a cylindrical lens, the diffraction patterns of the STOV beams in the focal plane of this cylindrical lens are captured using a CCD (BC106, Thorlabs). The 4 $f$ pulse shaper system form a STOV pulses generator, and the third grating and CCD form a STOV detector. The spatial intensity profiles of the STOV pulses in the focal plane of the spherical lens and the diffraction patterns of STOV pulses in the observed plane (OP) are measured or calculated.

## 2. Simulated results and experimental results

The wavepacket of an ideal STOV pulse can be expressed as [1]

$$E_1(x_1, y_1, t) = [t/b + i(l/|l|)y_1/a]^{|l|} E_0(x_1, y_1, t), \tag{S1}$$

where $E_0(x_1, y_1, w) = \exp[-(x_1^2 + y_1^2)/a^2]\exp[-t^2/b^2]$, $l$ is the topological charge. When $l = 0$, the STOV turns into a Gaussian pulse with waist radii of $a$ and $b$ in the spatial and temporal domain, respectively. The simulated STOV ($l = \pm 1, \pm 2,$) pulse profiles are shown in the top row of Fig. S1. The field in the spectrum domain can be calculated as

$$E_1(x_1, y_1, w) = \int_{-\infty}^{+\infty} E_1(x_1, y_1, t)\exp(-iwt)dt, \tag{S2}$$

then the STOV is diffracted by a grating, the field right after the grating can be written as [2]

$$E_2(x_2, y_2, w) = E_1(\beta x_2, y_2, w)\exp[i\gamma(w - w_c)x_2], \tag{S3}$$

where $\gamma = 2\pi/w_c d\cos(\theta_{d0})$, $\beta = \cos(\theta_{i0})/\cos(\theta_{d0})$, $\theta_{d0}$ and $\theta_{i0}$ are the diffracted angle and incident angle of the central frequency ($w_c$) ray, respectively. And the field propagates to the cylindrical lens in free space, the field before the cylindrical lens can be obtained as

$$E_3(x_3, y_3, w) = IFT\{FT\{E_2(x_2, y_2, w)\} \cdot H(f_x, f_y, w)\}, \tag{S4}$$

where $H(f_x, f_y, w) = \exp[ikz - i\pi\lambda z(f_x^2 + f_y^2)]$, and $k$ is the wave vector, *IFT* and *FT* denote the spatial inverse Fourier transform and Fourier transform. $f_x, f_y$ are spatial frequencies. Then the field pass through the cylindrical lens and can be expressed as

$$E_4(x_4, y_4, w) = E_3(x_3, y_3, w) \exp[-ikx_3^2/(2f)], \tag{S5}$$

$E_4(x_4, y_4, w)$ then propagates to the focal plane of the cylindrical lens. The intensity profile of the STOV and the diffraction pattern of STOV in the OP can be calculated as

$$I(x, y, w) = |E(x, y, w)|^2. \tag{S6}$$

The simulated intensity profiles and simulated diffraction patterns of STOV with $l$ = 1, 2, 3 4 and $l$ = -1, -2 ,-3, -4 are shown in row 1 and row 2 in Fig. S2(a) and (b), respectively. The diffraction patterns of STOVs show multi-lobe structures. For STOV with $l$ = n, the diffraction pattern has n + 1 lobes, and the number of gaps between the lobes is n. The measured intensity profiles and diffraction patterns of the corresponding STOVs are shown in row 3 and row 4 in Fig. S2(a) and (b), respectively. The feature structure of the diffraction pattern, such as the number of lobes and gaps are the same as the simulated results. However, there are some special structure of the experimental results, such as there are weak trails on the intensity profiles and the diffraction patterns, which are more obvious for higher order STOVs. The trails mainly result from the STOV pulses generated from the 4 *f* pulse shaper are not ideal STOVs [1, 3, 4]. The simulated STOV pulses generated from 4 *f* pulse shaper are shown in the bottom row of Fig. S1. As it can be seen that the STOVs generated from pulse shaper are a little bit elliptic and the energy distributions are not as uniform as the ideal STOVs. Furthermore, for higher order STOVs, there are n intensity nulls in the wavepackets which is different from those of the ideal STOVs, which have only one intensity nulls. The simulated results of the intensity profiles and diffraction patterns of STOVs generated using 4 *f* pulse shaper are shown in row 5 and row 6 in Fig. S2(a) and (b), respectively. The feature structures of the intensity profiles and diffraction patterns are consistent with those of the ideal STOVs. Moreover, as for the STOVs generated from 4 *f* pulse shaper, weak trails occurred on the two side of the patterns in the *y* direction. And the two head lobes show a little bit curve. This feature will be more obvious for STOVs with a relative broader spectrum as shown in Fig. S3. The features of the simulated STOV pulses generated from the 4 *f* pulse shaper get more closed to the measured results than those of the ideal STOV pulses.

The experimental results of STOVs with relative broad spectral width, a FWHM of about 20 nm, which can support a Fourier-transform-limited pulse duration of about 40 fs at central wavelength of 760 nm, are shown in Fig. S3. The feature structures, such as multi-lobe structure and the gap numbers are consistent with those of the corresponding STOV pulses with narrow spectral bandwidths discussed above. Since the STOV pulses are from the 4 *f* pulse shaper system, the weak trails appear and the two head lobes show a little bit of curve. The feature structures match the simulated results for the narrow band STOV pulses well.

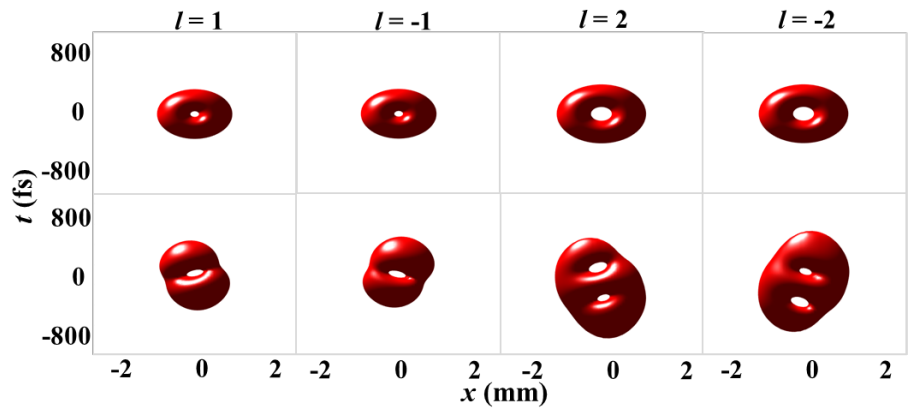

Figure S1. The simulated pulse profiles of the ideal STOVs ($l = \pm 1, \pm 2$,) (top row) and the simulated STOVs ($l = \pm 1, \pm 2$,) generated from 4$f$ pulse shaper (bottom row).

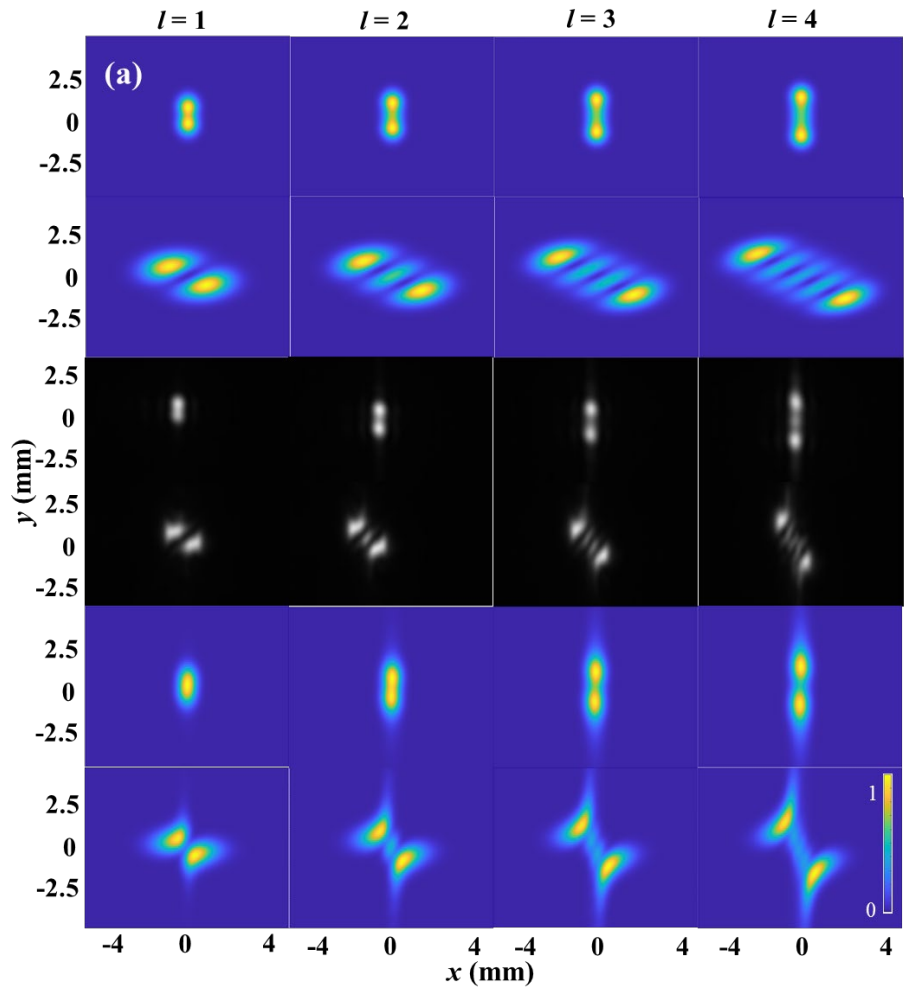

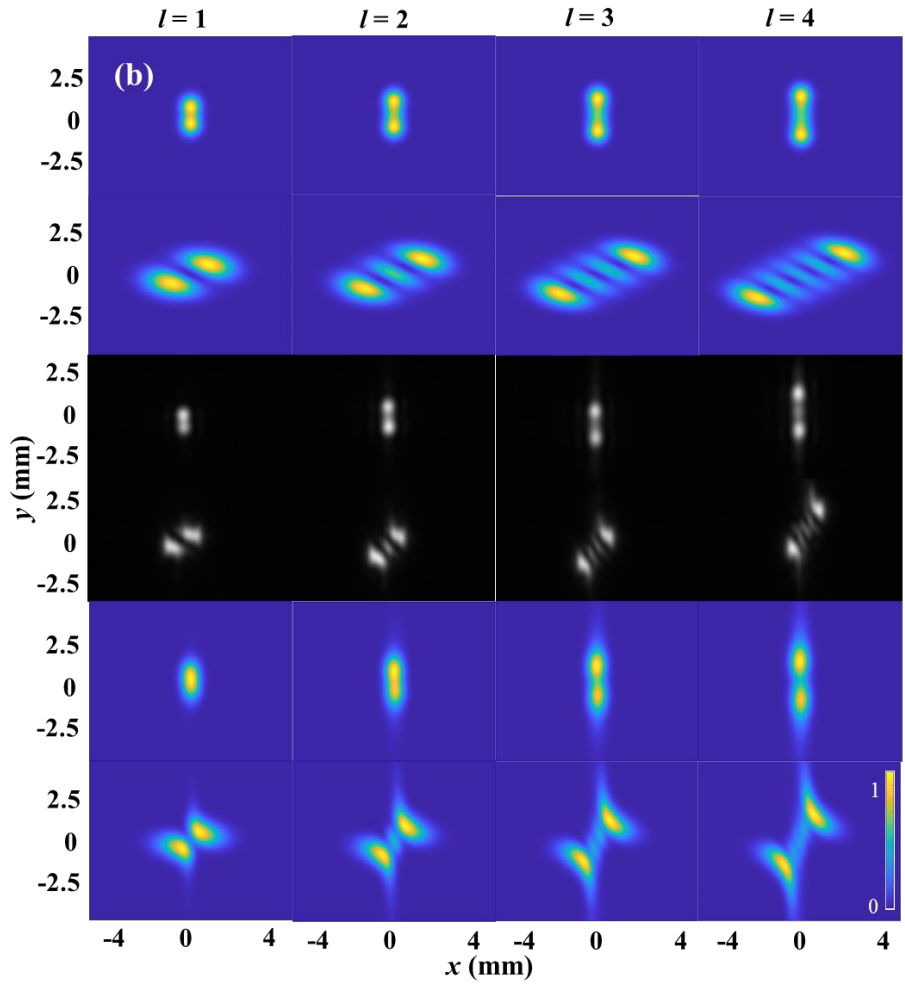

Figure S2. The simulated and measured intensity profiles and diffraction patterns of STOVs with $l = \pm 1, \pm 2, \pm 3, \pm 4$. (a) for STOV with positive topological, (b) for STOV with negative topological. row 1, 2, 3, 4, 5, 6 show the simulated intensity profiles and diffraction patterns for ideal STOVs, measured results for STOVs generated from $4f$ pulse shaper, and simulated results for STOVs generated from $4f$ pulse shaper.

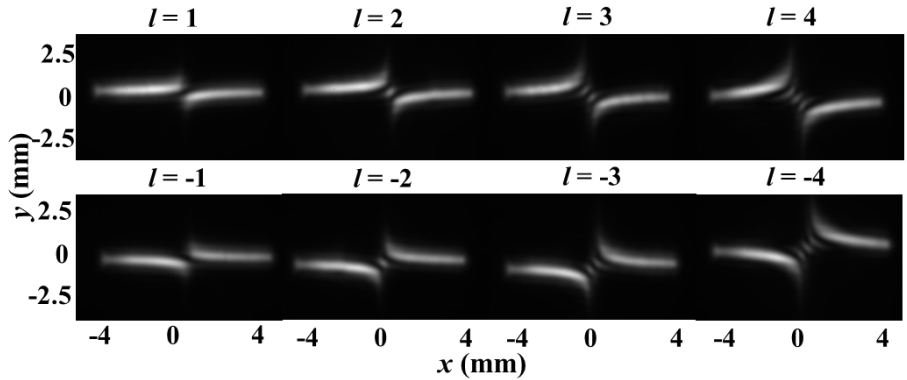

Figure S3. The measured diffraction patterns of STOVs ($l = \pm 1, \pm 2, \pm 3, \pm 4$) with spectral bandwidth of about 20 nm.